# General bound states in the continuum in momentum space

Qiao Jiang[1,2], Peng Hu[1], Jun Wang[1], Dezhuan Han[1,*], and Jian Zi[3,†]

[1]*College of Physics, Chongqing University, Chongqing 401331, China.*
[2]*Chongqing Key Laboratory for Strongly Coupled Physics, Chongqing University, Chongqing 401331, China.*
[3]*Department of Physics, Key Laboratory of Micro- and Nano-Photonic Structures (MOE), and State Key Laboratory of Surface Physics, Fudan University, Shanghai 200433, China.*

*\*e-mail*: dzhan@cqu.edu.cn
[†]*e-mail*: jzi@fudan.edu.cn

## Abstract

Polarization singularities including bound states in the continuum (BICs) and circularly polarized states have provided promising opportunities in the manipulation of light waves. Previous studies show that BICs in photonic crystal slabs are protected by $C_2T$ symmetry and hence normally exist on the high-symmetry lines of momentum space. Here, we propose an approach based on graph theory to study these polarization singularities in momentum space, especially in the region off the high-symmetry lines. With a polarization graph, it is demonstrated for the first time that BICs can stably exist off the high-symmetry lines of momentum space for both one-dimensional and two-dimensional photonic crystal slabs. Furthermore, two kinds of interesting processes, including the merging involved with this newly found BICs both on and off the high-symmetry lines, are observed by merely changing the thickness of photonic crystal slab. Our findings provide a new perspective to explore polarization singularities in momentum space and render their further applications in light-matter interaction and light manipulation.

## Introduction

Photonic crystal (PhC) slabs that support resonance modes have become a platform for studying singular optics [1-3] in momentum space [4-7], where different kinds of polarization singularities, including vortex polarization singularities (V points) and circularly polarized states (C points), are revealed and experimentally observed. At a V point, as the far-field radiation vanishes and the corresponding Bloch mode has no leakage, it is dubbed the bound state in the continuum (BIC) [4-12]. The unique properties of BICs, including the polarization vortex and extremely high Q factor, have provided promising applications in nonlinear optics [13-17], lasers [18-22] and light field manipulation [23, 24]. On the other hand, a C point with finite Q factor can only couple to circularly polarized waves, leading to possible applications of chiral light sources [25].

Besides the above interesting properties, the topological nature of these singularities has triggered a further study of the evolution of these singularities in momentum space. Due to the robustness of BICs, they can be continuously moved along the high-symmetry lines of momentum space by varying the geometrical parameters while keeping the symmetry of PhC slab. This has been used to generate ultrahigh-Q guided resonances by merging different BICs [26]. If the symmetry of the PhC slab is broken, more interesting phenomena of these polarization singularities have been discovered. For example, a pair of C points with the same topological charge but opposite handedness can be generated from a BIC by breaking the $C_2$ symmetry of the PhC slab [6]. When further breaking the up-down mirror symmetry, two C points spilt from an off-Γ BIC has been utilized to generate a unidirectional guided resonance by recombining them, which can only radiate to one side of the PhC slab [27]. It is worth noting that although C points can exist and evolve off the high-symmetry lines of momentum space, the initial or recombined V points in the unidirectional guided resonance are always on the high-symmetry lines [27, 28]. Recently, it is reported that BICs off the high-symmetry lines can be found by introduce PT-symmetric perturbation [29] or be tuned from high-symmetry lines by breaking all in-plane mirror symmetries [30]. However, it is still a puzzle whether there is a BIC off the high-symmetry lines of momentum space which exists intrinsically in the system of PhC slabs without changing its symmetry. This is crucial for further understanding the topological nature of the leaky modes in momentum space.

In this work, we study the distribution and evolution of polarization singularities through the polarization graph in momentum space. The vertices and edges of the graph correspond to the polarization singularities and nodal lines. We demonstrate that the topological charge of a face in the polarization graph is exactly zero, which indicates that the topological charges of the singularities at different $k$ points are in fact related to each other. Thus, it reveals a nonlocal property of polarization singularities in momentum space. Based on this property of polarization graph, the existence of BICs off the high-symmetry lines of momentum space without reducing the system symmetry is demonstrated for the first time, which are called *general* BICs here. Furthermore, by merely changing the thickness of the PhC slab, this new type of BIC is demonstrated to be robust, and two kinds of interesting processes related to this BIC are observed, including its merging with an ordinary off-Γ BIC on the high-symmetry lines and the merging together with another general BIC off the high-symmetry lines of momentum space.

**Results**

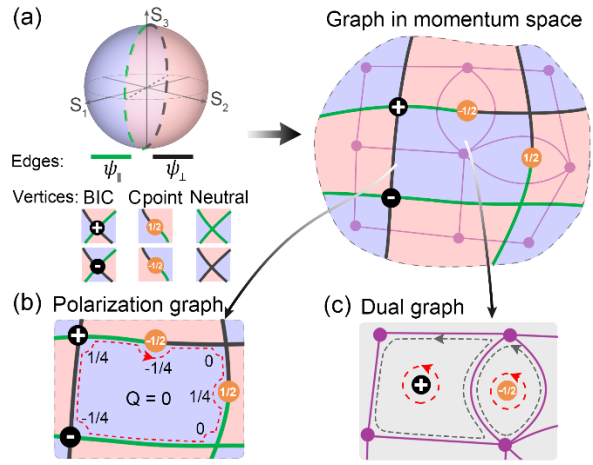

**Fig. 1** Polarization graph in momentum space. (a) Schematic diagram of a polarization graph and its dual graph. In the polarization graph, the edges correspond to polarization nodal lines of major axis of polarization ellipse (green lines: $\psi_\parallel$, black lines: $\psi_\perp = \psi_\parallel + \pi/2$). BICs and C points correspond to the intersection and joint points of edges with different colors, respectively. The dual graph is shown by the purple vertices and edges. (b) The charge of a bounded face in polarization graph is always zero. Numbers near the red dashed contour indicate shared charges of the corresponding vertices to this face. (c) The charge of a bounded face in the dual graph is exactly the charge of the corresponding vertex in the original graph.

In general, the far-field polarization states for resonance states are elliptical and transverse to the wave vector ***k***, which can be projected onto the *x-y* plane in an appropriate way with their polarization information preserved [7]. In this projected polarization field, polarization singularities, including both BICs (V points) and C points, correspond to vortex centers with undefined orientation angle of the major axis for the polarization ellipse. As illustrated in Fig. 1(a), to construct a polarization graph that can reveal the correlation between different polarization singularities in momentum space, the edges are defined as polarization nodal lines with orthogonal orientation angle $\psi_\parallel$ (green lines) and $\psi_\perp$ (black lines). These edges correspond to polarization states on two longitudes on a great circle on Poincaré sphere, dividing the sphere into two hemispheres (colored as red and blue) [see left panel of Fig. 1(a)]. The polarization states of different faces of a polarization graph can be classified as red and blue accordingly. With this definition, there are three typical types of vertices: BICs, C and neutral points. BICs and C points correspond to the intersection and joint points of different kinds of polarization nodal lines (i.e. edges with different colors), which can be characterized by a topological charge $q$ as:

$$q = \frac{1}{2\pi} \oint_L d\boldsymbol{k}_\parallel \cdot \nabla_{\boldsymbol{k}_\parallel} \psi(\boldsymbol{k}_\parallel), \tag{1}$$

where $L$ is a counterclockwise closed loop around a polarization singularity in momentum space. $\boldsymbol{k}_\parallel = (k_x, k_y)$ is the in-plane wave vector. $\psi(\boldsymbol{k}_\parallel) = \frac{1}{2}\arg(S_1 + iS_2)$, where $S_i$ is the $i$-th Stokes parameter, is the orientation angle of the major axis for the polarization ellipse. The topological charge for a BIC and a C point is an integer and a half integer [2-4, 6], respectively. Different from BICs and C points, neutral points are formed by two edges with the same color and carry zero topological charge.

For illustration, considering a general case involving both BICs and C points in momentum space, a polarization graph and its dual graph (colored in purple) are shown in the right panel of Fig. 1(a). The dual graph has a vertex corresponding to each face of the polarization graph, and a face for each vertex in the polarization graph. It can be seen that different polarization singularities are correlated through edges and a series of faces is formed in the polarization graph. Note that a face can be unbounded when it extends beyond the light line or to infinity (e.g. when $k\to\infty$ along the direction that is uniform for a one-dimensional PhC slab). Here, we focus on the faces bounded by edges in the polarization graph. As shown in Fig. 1(b), the topological charge $Q$ of a bounded face is defined as the winding number along a clockwise contour $C$ (red dashed line) inside this face. Because the polarization along the edges remains its major axis in the same direction and thus does not contribute to the winding number, the result is only determined by the integral along the small arcs around the polarization singularities. For an arc, since the orientation of major axes at the starting and end points is orthogonal, its contribution to winding number of the face is $q/n$, where $n$ is the degree of this vertex, i.e. the number of edges that are incident to it (see Sec. S1 of Supplemental Material). We define $q_i$ as the shared topological charge of the $i$-th polarization singularity to this face. On the other hand, since there are no singularities inside this contour, the total winding number should be zero. Thus, we obtain the following relation:

$$Q = \frac{1}{2\pi} \oint_C d\boldsymbol{k}_\parallel \cdot \nabla_{\boldsymbol{k}_\parallel} \psi(\boldsymbol{k}_\parallel) = \sum q_i = 0, \quad (2)$$

which means the topological charge $Q$ of an arbitrary bounded face in polarization graph equals to the sum of $q_i$ and should be zero. Similarly, as shown in Fig. 1(c), a topological charge calculated along the counterclockwise contour (gray dashed line) can also be assigned to a face of the dual graph. Since the face in the dual graph corresponds to a vertex in the polarization graph, this topological charge of a face is actually the charge of a vertex in the polarization graph, equaling to the result calculated by Eq. (1) along a closed loop [red dashed lines in Fig. 1(c)]. In fact, if we let a flux $\boldsymbol{B} = \frac{1}{2\pi} \nabla_{\boldsymbol{k}_\parallel} \psi(\boldsymbol{k}_\parallel)$, Eqs. (1) and (2) can be regarded as "*Ampère's circuital law*" for a polarization graph in momentum space, which provides a unified description for the charges of faces and vertices in a polarization graph. In particular, Eq. (2) reveals a remarkable correlation of different polarization singularities and is useful for exploring the existence and evolution of the singularities in momentum space.

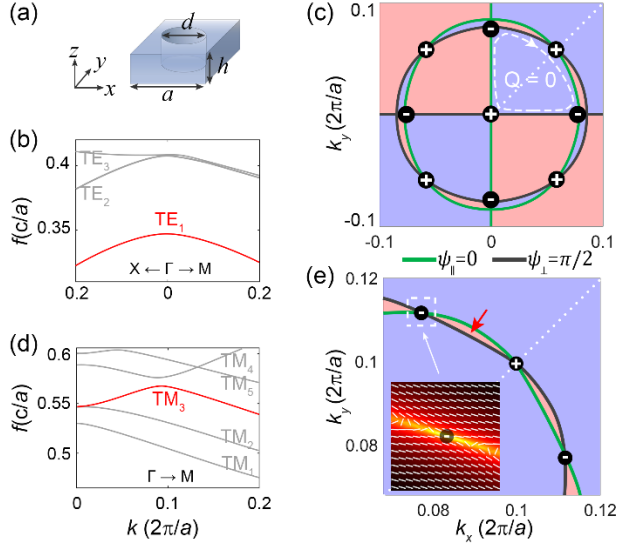

**Fig. 2** Polarization graph and general BICs in a PhC slab. (a) Schematic of a two-dimensional PhC slab consisting of a square lattice of circular air holes in a dielectric slab. (b) Simulated TE-like band structure along the X-Γ-M direction for a PhC slab studied in Ref. [26]. (c) Polarization graph of $TE_1$ band [red line in (b)], exhibiting one at-Γ BIC and eight off-Γ BICs. (d) Simulated TM-like band structure along the Γ-M direction. (e) Polarization graph of $TM_3$ band. The inset shows the far-field polarization states and Q factor near the general BIC at $\boldsymbol{k}_\parallel = (0.0775, 0.1115)2\pi/a$. The white dotted line in (c) and (e) indicates the Γ-M direction. The blue and red colors in (c) and (e) indicate the regions with Stokes parameter $S_2<0$ and $S_2>0$, respectively.

Here, we first utilize polarization graph to investigate BICs supported by a two-dimensional PhC slab reported in Ref. [26]. The PhC slab is formed by square-lattice circular air holes in a dielectric slab [see Fig. 2(a)]. For the lowest TE-like band [$TE_1$ in Fig.2 (b)], there are nine BICs, including one at-Γ BIC and eight off-Γ BICs. It has been experimentally shown that these BICs can be tuned together to generate a merging BIC with ultrahigh Q factor. However, it is still unclear why the topological charges for the eight off-Γ BICs should alternate between +1 and -1. In Fig. 2(c), we plot the polarization graph of $TE_1$ band which is obtained from the simulated far-field polarization in momentum space. Considering the in-plane mirror symmetry ($\sigma_x$ and $\sigma_y$) of the structure, the polarization nodal lines of $\psi_\parallel = 0$ (green lines) and $\psi_\perp = \pi/2$ (black lines) are chosen. It can be observed that these BICs indeed locate at the vertices connected by different edges. Interestingly, as indicated by the white dashed contour, there must be two BICs with charge of +1 and the other two with charge of -1 since the net charge of the bounded face is zero according to Eq. (2). Due to the $C_{4v}$ symmetry of the PhC slab, the charges of off-$\Gamma$ BICs on the $k_x$ and $k_y$ axes are the same. Thus, when we know the topological charge of at-$\Gamma$ BIC is +1, it can be directly determined that the BIC on $\Gamma$-M direction have a charge of +1 and another two BICs on the the $k_x$ and $k_y$ axes should possess a charge of -1. This example suggest that polarization graph is a powerful tool for understanding the charge distribution in momentum space.

It has been widely demonstrated that BICs can robustly exist on the high-symmetry lines of momentum space in a PhC slab with symmetry of $C_2^z T$ and $\sigma_z$. Polarization graph further provides a way to find undiscovered BICs that exist off the high-symmetry lines, namely, the general BICs. To verify this, we still adopt a square-lattice PhC slab as schematically shown in Fig. 2(a); the dielectric slab consists of $Si_3N_4$ (refractive index $n = 2$) and period $a = 400$ nm, hole diameter $d = 0.5a$, and thickness $h = 1.6a$. The band under study is $TM_3$ as highlighted in red in Fig. 2(d). Figure 2(e) plots the polarization graph near a BIC (with topological charge of +1) on $\Gamma$-M direction. It exhibits another two BICs at $\boldsymbol{k}_\| = (0.0775, 0.1115)2\pi/a$ and $\boldsymbol{k}_\| = (0.1115, 0.0775)2\pi/a$, which are symmetrical with respect to the $\Gamma$-M direction due to the mirror symmetry. Focusing on the BIC on the upper left (indicated by the white dashed box), its topological charge should be -1 because it forms a bounded face together with the BIC on $\Gamma$-M direction as indicated by the red arrow. Again, the zero net charge of a face shown in Eq. (2) can be applied to determine the charge. For confirmation, the polarization states and Q factors are shown in the inset, where we can see a V point with a topological charge of -1 as well as a divergent Q factor.

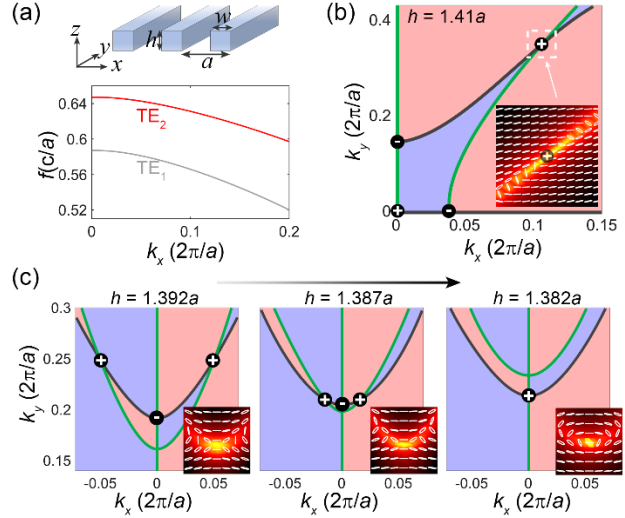

**Fig. 3** Merging of two general BICs and an off-Γ BIC. (a) Schematic diagram of a one-dimensional PhC slab and simulated TE-like band structure for $h = 1.41a$. (b) Simulated polarization graph for the TE$_2$ band. The inset shows far-field polarization states and Q factors near the general BIC (indicated by the white dashed box). (c) Simulated polarization graphs for $h = 1.392a$, $1.387a$, and $1.382a$. The insets show the polarization states and Q factors near off-Γ BIC on the $k_y$ axis.

It is worth noting that the general BIC uncovered above exists intrinsically off the high-symmetry lines of momentum space, distinct from the BICs tuned from the high-symmetry lines by breaking the in-plane mirror symmetries [30], which can only be realized in a two-dimensional PhC slab. For further verification, we exemplify a one-dimensional Si$_3$N$_4$ PhC slab with period $a$ = 400 nm, width $w = 0.58a$, and thickness $h = 1.41a$ [see Fig. 3(a)]. The band under study is TE$_2$. As shown in Fig. 3(b), due to the in-plane mirror symmetry of this PhC slab, the polarization graph for a quarter of momentum space is plotted to exhibit the distribution of BICs [see more in Sec. S2 of Supplemental Material]. It can be observed that there are four BICs, including three ones on the high-symmetry lines ($k_x$ and $k_y$ axes) and one off the high-symmetry lines. The first three BICs are easy to find by simulations along the high-symmetry lines [4, 31], but the general one has never been seen before. The polarization states and Q factors near the general BIC are plotted in the inset of Fig. 3(b), which clearly demonstrate its existence with a divergent Q factor and a topological charge of +1.

When we continuously vary the geometric parameters of this PhC slab without breaking the symmetry, these BICs will evolve with the polarization graph and robustly exist until they collide with each other. Figure 3(c) shows the polarization graphs for $h = 1.392a$, $1.387a$, and $1.382a$, which indicates that two general BICs (charge +1) gradually approach the off-$\Gamma$ BIC (charge -1) on $k_y$ axis when decreasing the thickness. Due to the left-right mirror symmetry ($\sigma_x$), these three BICs will merge on the $k_y$ axis. After the merging process, only an off-$\Gamma$ BIC with charge +1 remains, obeying the conservation law of topological charge. It is interesting that the topological charge of off-$\Gamma$ BIC can be changed by merely varying the geometric parameters of this PhC slab. This can be further demonstrated by the polarization states and Q factors in the insets of Fig. 3(c), which shows that the topological charge of the off-$\Gamma$ BIC changes from -1 to +1. Note that a similar process can also happen for the case of two-dimensional PhC slab in Fig. 2(e) [see Sec. S3 of Supplemental Material].

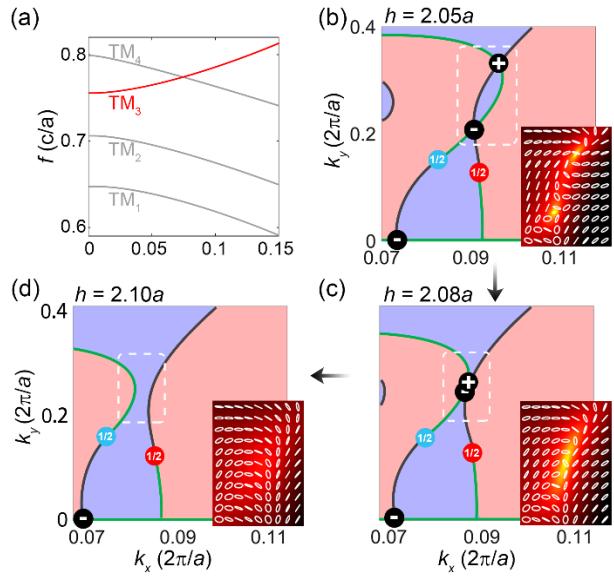

**Fig. 4** Merging two general BICs off the high-symmetry lines. (a) Simulated TM band structure of a one-dimensional $Si_3N_4$ PhC slab with $w = 0.47a$ and $h = 2.05a$. (b)-(d) Polarization graphs for (b) $h = 2.05a$, (c) $2.08a$, and (d) $2.10a$. The insets are far-field polarization states and the corresponding Q factors near the two general BICs (marked by the white dashed box). The blue (red) dot indicates right-handed (left-handed) C point.

The above examples show that general BICs can merge with off-$\Gamma$ BIC on the high-symmetry lines of momentum space. Here, we further demonstrate that two general BICs can also merge off the high-symmetry lines in a one-dimensional PhC slab. The simulated TM-like band structure for $h = 2.05a$ is plotted in Fig. 4(a). As shown in Fig. 4(b), the polarization graph of band $TM_3$ exhibits various polarization singularities, including two C points and three BICs [see more details in Sec. S4 of Supplemental Material]. It is obvious that an arbitrary bounded face in Fig. 4(b) carries zero net charge as we demonstrated in Eq. (2). Focusing on the two general BICs, the polarization graph in Fig. 4(c) shows that they will merge together when $h$ is increased to $2.08a$, which can be clearly observed from the polarization states and Q factors. When further increasing $h$, these two BICs annihilate each other and the Q factor is no longer divergent [see Fig. 4(d)]. This process can be explicitly observed in the evolution of the polarization graph, which shows a typical configuration from crossing to separation of two different edges when increasing $h$ [see the white dashed box in Figs. 4(b)-4(d)].

**Conclusions**

In summary, we have studied the distribution and evolution of polarization singularities in momentum space based on the graph theory of polarization states. The topological charge for an arbitrary bounded face in the polarization graph is demonstrated to be zero, manifesting the correlation of polarization singularities at different $k$ points. This result manifest a nonlocal property of singularities and is an extension of the conversation law of topological charges. It plays a significant role in exploring the topological charges in the whole momentum space, especially off the high-symmetry lines. With the polarization graph, it is firstly demonstrated that BICs can actually exist off the high-symmetry lines of momentum space without breaking the mirror symmetry of the PhC slab. By only varying the thickness of the PhC slab, it is revealed that the general BICs can merge with an ordinary off-$\Gamma$ BIC on the high-symmetry lines and merge together off the high-symmetry lines of momentum space. Our findings complete the classification and dynamics of polarization singularities in momentum space instead of being limited to the vicinity of high-symmetry lines, and possibly inspire further advances in light manipulation and light-matter interaction.

**Acknowledgements**

We thank Professors C. M. Song, C. Peng, L. Lu, and L. Shi for helpful discussions. This work is supported by the National Natural Science Foundation of China (Grant Nos. 12047564, 12074049, 12104073, and 12147102), China Postdoctoral Science Foundation (Grant No. 2020M683234), Fundamental Research Funds for the Central Universities (Grant No. 2022CDJQY-007).

Q. J. and P. H. contributed equally to this work.

# Supplemental Material

# General bound states in the continuum in momentum space

Qiao Jiang[1,2], Peng Hu[1], Jun Wang[1], Dezhuan Han[1,2], and Jian Zi[3]

[1]*College of Physics, Chongqing University, Chongqing 401331, China.*
[2]*Chongqing Key Laboratory for Strongly Coupled Physics, Chongqing University, Chongqing 401331, China.*
[3]*Department of Physics, Key Laboratory of Micro- and Nano-Photonic Structures (MOE), and State Key Laboratory of Surface Physics, Fudan University, Shanghai 200433, China.*

## S1. Illustration of shared charge in the polarization graph

As the far-field polarization state with orientation angle $\psi$ and ellipticity angle $\chi$ can be mapped to a point on Poincaré sphere with longitude and latitude coordinates of $(2\psi, 2\chi)$, a closed loop L in momentum space that encloses a polarization singularity with topological charge $q$ will correspond to a closed trajectory on the Poincaré sphere. The winding number of this trajectory on the Poincaré sphere around the $S_3$ axis is $n_w = 2q$. For example, the trajectory for a closed counterclockwise loop around a BIC with $q = 1$ goes around the $S_3$ axis two times in the counterclockwise direction. With this correspondence, it is intuitive to understand the charge sharing in the polarization graph.

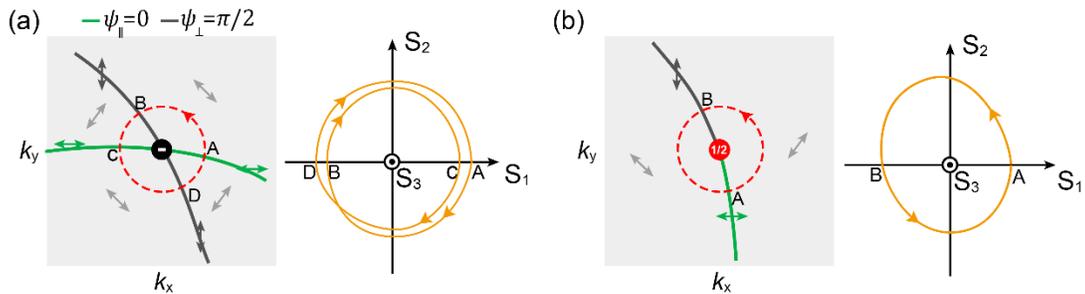

**Fig. S1.** (a) Schematic of polarization graph near a BIC with topological charge of -1(left panel)

and the corresponding trajectory in the $S_1$-$S_2$ plane for a closed loop shown by the red dashed circle around the BIC (right panel). (b) Schematic of polarization graph for a C point with topological charge of 1/2 and the corresponding projected trajectory in the $S_1$-$S_2$ plane for a closed loop around the C point.

Here, we take a BIC with charge $q_{BIC} = -1$ and a C point with charge $q_{CP} = \frac{1}{2}$ for illustration. As shown in Fig. S1(a), for a BIC with charge $q_{BIC} = -1$, the closed loop (red dashed circle) in momentum space is made up of four parts: $\widehat{AB}$, $\widehat{BC}$, $\widehat{CD}$, and $\widehat{DA}$. When mapped onto Poincaré sphere, the corresponding projected trajectory in the $S_1$-$S_2$ plane is illustrated in the right panel of Fig. S1(a). Focusing on the path AB in momentum space [see left panel of Fig. S1(a)], because the polarization nodal line of $\psi_\parallel = 0$ ($\psi_\perp = \pi/2$) corresponds to $S_2=0$ and $S_1>0$ ($S_1<0$), the corresponding trajectory in the $S_1$-$S_2$ plane will go around $S_3$ axis with an angle of $-\pi$, exactly resulting in a shared topological charge $q_{AB} = -1/4 = \frac{1}{4} q_{BIC}$. The same also holds for the paths BC, CD, and DA, that is, $q_{AB} = q_{BC} = q_{CD} = q_{DA} = \frac{1}{4} q_{BIC}$. For a C point with charge $q_{CP} = 1/2$, as shown in Fig. S1(b), the closed loop (red dashed circle) in momentum space is made up of $\widehat{AB}$ and $\widehat{BA}$, corresponding to a closed trajectory around the $S_3$ axis with winding number $n_w = 1$ [see right panel of Fig. S1(b)]. Similar to the case of BIC in Fig. S1(a), the path AB in momentum space corresponds to a trajectory going around $S_3$ axis with an angle of $\pi$ in the $S_1$-$S_2$ plane, indicating a shared topological charge $q_{AB} = 1/4 = \frac{1}{2} q_{CP}$. Similarly, we have $q_{AB} = q_{BA} = \frac{1}{2} q_{CP}$.

## S2. Details of polarization graphs for the examples in Fig. 3

In Fig. 3(b) of the main text, we only plot the polarization graph for a quarter of momentum space due to the in-plane mirror symmetry of the one-dimensional PhC slab. Fig. S2(a) provides a whole view of the polarization graph and Q factor distribution for $h = 1.41a$. It can be clearly seen that there are four general BICs distributed symmetrically about $k_x$ and $k_y$ axes.

When gradually decreasing the thickness of the PhC slab, there is a process of BIC merging at Γ point before the merging of two general BICs and an off-Γ BIC shown in Fig. 3(c). As shown in Fig. S2(a) and S2(c), two off-Γ BICs on the $k_x$ axis move towards the at-Γ BIC as $h$ gradually decreases. After a merging process, the topological charge of at-Γ BIC changes from +1 to -1. When further decreasing the thickness, the process shown in Fig. 3(c) will happen.

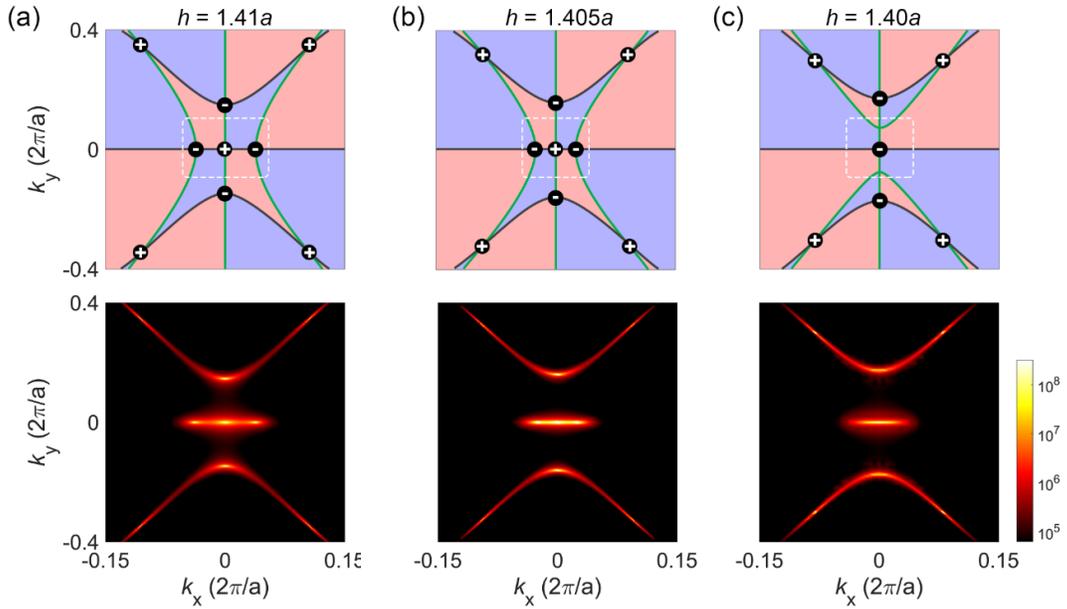

**Fig. S2** Polarization graphs and Q factors of TE$_2$ band in Fig. 3 for (a) $h = 1.41a$, (b) $1.41a$, and (c) $1.40a$.

S3. **Merging of general BICs and an off-Γ BIC in a two-dimensional PhC slab**

Here, we show that the two general BICs in Fig. 2(e) can merge with the off-Γ BIC in the Γ-M direction when decreasing the thickness of the PhC slab. For comparison, polarization graphs and corresponding Q factors for $h = 1.60a$ and $1.585a$ are plotted in Figs. S3(a) - S3(d). It can be observed that only the off-Γ BIC remains when *h* is decreased from 1.60*a* to 1.585*a*. Due to the conservation of topological charge, charge of this BIC will change from +1 to -1, which can be clearly demonstrated by the polarization states in Figs. S3(e) and S3(f).

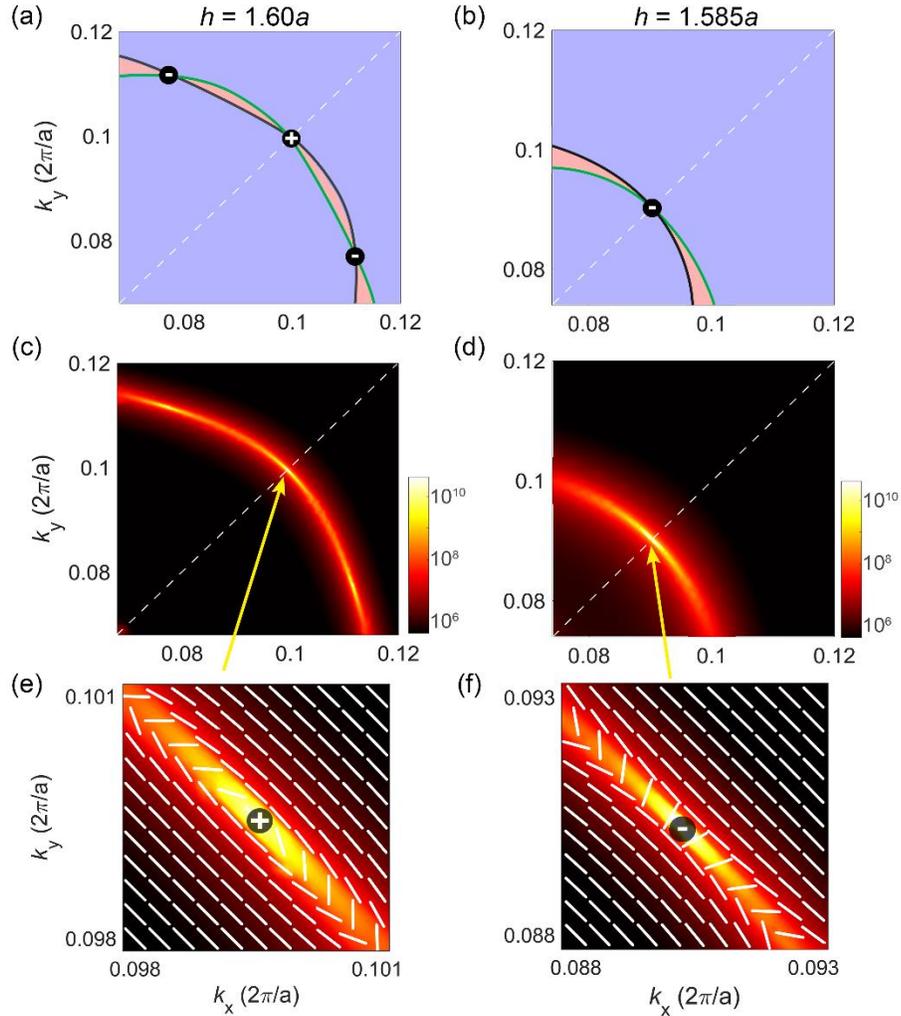

**Fig. S3** (a, b) Polarization graphs for *h* = (a) 1.60*a* and (b) 1.585*a*. (c, d) Q factor profiles for *h* = (c) 1.60*a* and (d) 1.585*a*. (e, f) Polarization states and Q factor near the off-Γ BIC in the Γ-M direction for *h* = (e) 1.60*a* and (f) 1.585*a*. The white dashed lines indicate the high-symmetry lines along the Γ-M direction.

## S4. Details of polarization states for band TM$_3$

In Fig. 4 of the main text, we focus on the two general BICs off the high-symmetry lines of momentum space. Here, to provide a clear proof of the other polarization singularities shown in Fig. 4(b), the detailed polarization states are plotted in Fig. S4(a) for $h = 2.05a$. Due to the symmetry of the PhC slab, we only plot the calculated results for $k_y \geq 0$. The right-handed (left-handed) C point are marked by blue (red) dots. From the polarization states, the topological charges of the off-$\Gamma$ BIC on the $k_x$ axis and two C points can be obtained as -1 and 1/2, respectively. The divergent points of the Q factor in Fig. S4(b) clearly demonstrate the existence of two general BICs and an off-$\Gamma$ BIC in this one-dimensional PhC slab.

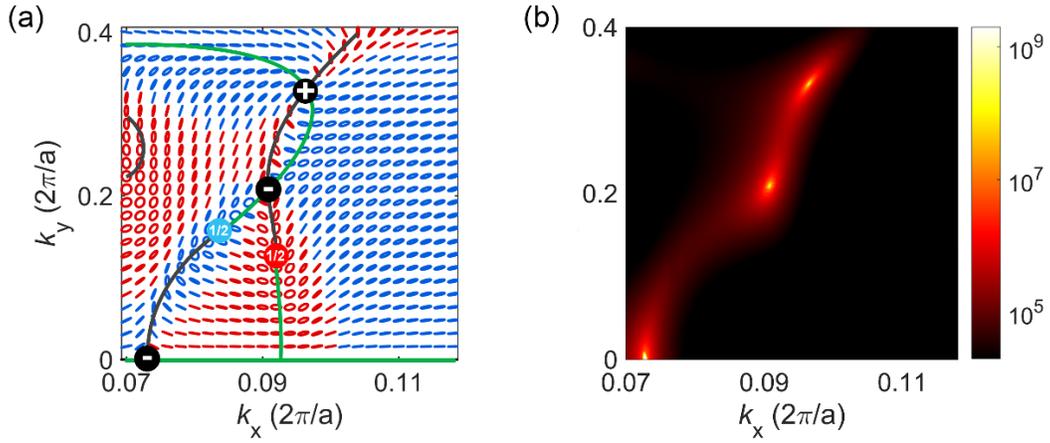

**Fig. S4** (a) Polarization states and (b) Q factor profile of $TM_3$ band corresponding to the polarization graph in Fig. 4(b). The polarization nodal lines in (a) are the same with Fig. 4(b). The blue (red) ellipse in (a) denotes the right-handed (left-handed) elliptic polarization.